\documentclass[11pt]{article}

    \usepackage[preprint]{neurips_2023}

\usepackage[utf8]{inputenc} 
\usepackage[T1]{fontenc}    
\usepackage{hyperref}       
\usepackage{url}            
\usepackage{booktabs}       
\usepackage{amsmath}
\usepackage{amsfonts}       
\usepackage{nicefrac}       
\usepackage{microtype}      
\usepackage{xcolor}         

\usepackage{graphicx}

\usepackage{amsbsy}
\usepackage{amsmath}
\usepackage{amssymb}
\usepackage{mathtools}
\usepackage{multirow}
\usepackage{multicol}
\usepackage{wrapfig}

\usepackage{adjustbox}

\newcommand{\mname}{Wave-LSTM}

\NewDocumentCommand{\rot}{O{30} O{3em} m}{\makebox[#2][l]{\rotatebox{#1}{#3}}}%
\NewDocumentCommand{\rott}{O{30} O{2em} m}{\makebox[#2][l]{\rotatebox{#1}{#3}}}%
\NewDocumentCommand{\rotS}{O{60} O{3em} m}{\makebox[#2][l]{\rotatebox{#1}{#3}}}%

\def\0{\mathbf{0}}

\def\j{{j}}
\def\coef{{\alpha}}

\usepackage{lipsum}
\usepackage{fancyhdr}       
\usepackage{graphicx}       
\graphicspath{{media/}}     
\usepackage{caption}
\usepackage{subcaption}

\title{Wave-LSTM: Multi-scale analysis of somatic whole genome copy number profiles}

\author{%
  Charles Gadd \\
  University of Oxford \\
  \And
  Christopher Yau \\
  University of Oxford \\
  Alan Turing Institute \\
  Health Data Research UK \\
}

\begin{document}

\maketitle

\begin{abstract}

Changes in the number of copies of certain parts of the genome, known as copy number alterations (CNAs), due to somatic mutation processes are a hallmark of many cancers. This genomic complexity is known to be associated with poorer outcomes for patients but describing its contribution in detail has been difficult. Copy number alterations can affect large regions spanning whole chromosomes or the entire genome itself but can also be localised to only small segments of the genome and no methods exist that allow this multi-scale nature to be quantified. In this paper, we address this using Wave-LSTM, a signal decomposition approach designed to capture the multi-scale structure of complex whole genome copy number profiles. Using wavelet-based source separation in combination with deep learning-based attention mechanisms. We show that Wave-LSTM can be used to derive multi-scale representations from copy number profiles which can be used to decipher sub-clonal structures from single-cell copy number data and to improve survival prediction performance from patient tumour profiles.
\end{abstract}

\section{Introduction}
\begin{figure}[!t]
    \centering
\includegraphics[width=0.75\textwidth]{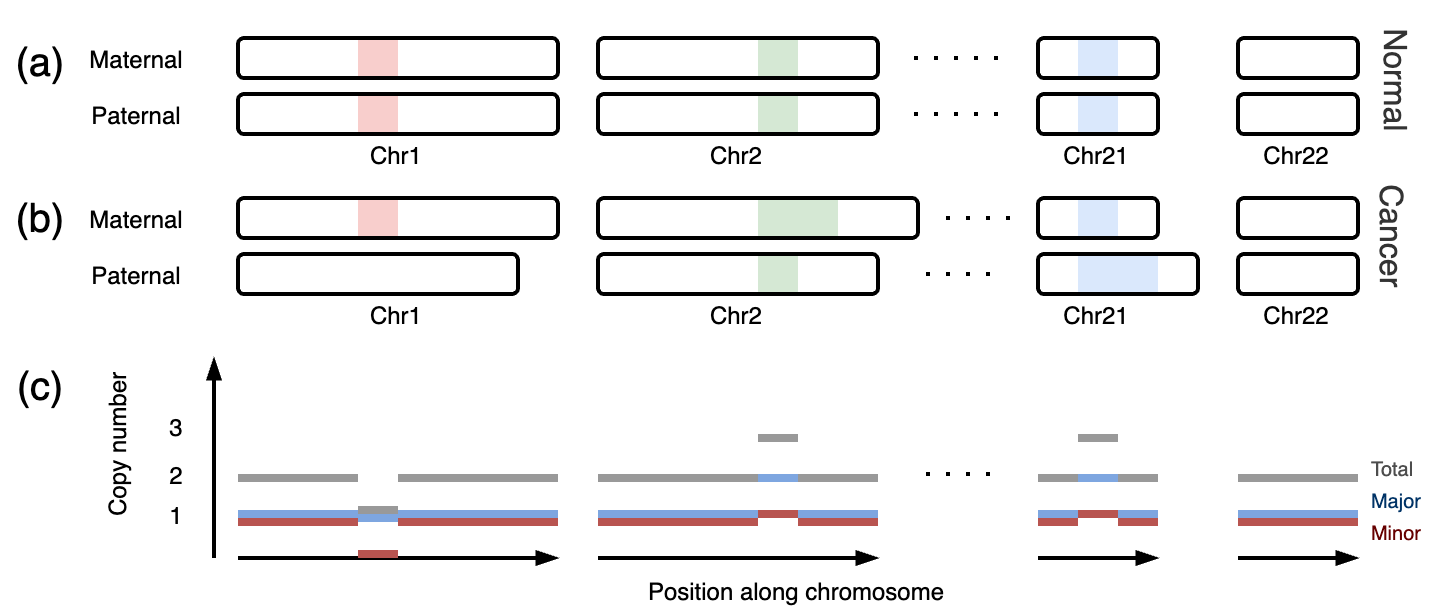}
    \caption{\textbf{An example of a somatic copy number alteration}. (\textbf{a}) A normal cell. (\textbf{b}) A cancerous cell which has undergone a deletion on the paternal strand of chromosome (chr) $1$, and an insertion on the maternal and paternal strands of chromosome $2$ and $21$, respectively. (\textbf{c}) These cancerous mutations are quantified by the copy number.}
    \label{fig:cna}
\end{figure}

Somatic copy number alterations (SCNAs) are a hallmark of cancers \citep{beroukhim2010landscape}. The normal human genome typically contains two copies of every gene (Figure \ref{fig:cna}a), one copy derived from each biological parent, and a normal complement of 44 non-sex chromosomes. In a cancer genome, errors in DNA repair and dysregulation of cellular processes due to mutations can alter the genomic content leading to the loss or gain of genomic segments (Figure \ref{fig:cna}b). This can lead to a phenomenon known as \emph{chromosomal instability} (CIN) in which large-scale whole genome changes occur \citep{sansregret2018determinants}. Copy number changes affecting certain genes can lead to an oncogenic effect in which additional copies of a gene (oncogene) promote cancer development or the loss of a gene (tumour suppressor gene) stops a suppressive role that normally occurs. 

Modern molecular analysis technologies, such as DNA sequencing, together with bioinformatics (such as TITAN \citep{ha2014titan} or ASCAT \citep{van2010allele}), allow the number of total copies of each part of the genome to be measured and counted. This work is not concerned with the low-level processing of raw sequencing data. Instead, we focused on the analysis of the derived copy number profiles from the processed sequenced data which consists of \emph{major} and \emph{minor} copy numbers or so-called \emph{unphased} copy number counts (Figure \ref{fig:cna}c). Therefore, the data we consider consists of the following: chromosome, position along chromosome, major copy number and minor copy number (note it is normally not  possible to resolve to the level of parentally derived copy number without special experimental techniques). There are typically millions of genomic positions measured.

Copy number alterations (CNAs) occupy multiple genomic scales. Some alterations affect the whole genome itself -- such as \emph{whole genome doubling} (WGD) \citep{bielski2018genome} -- where failed cellular division causes daughter cells to gain a whole extra copy of the genome. Other changes may only affect individual or partial chromosomes, or are localised to small parts of the genome. Evolutionary forces and biological constraints (i.e. cellular viability) often drive cancers from different individuals and types to acquire recurrent patterns of SCNAs. Recent studies have sought to understand their underlying aetiology \citep{drews2022pan} and have discovered associations with severity of disease and clinical outcomes~\citep{beroukhim2010landscape}. However, it remains a challenge to understand how the mixture of large and fine-scale CNAs lead to certain outcomes (e.g. survival).

{\bf Contributions.}~In this paper, we present an approach to learn multi-scale representations of high-dimensional, structured cancer copy number profiles called Wave-LSTM. We demonstrate through examples that Wave-LSTM provides a means for exploratory and predictive analyses that enable us to understand the multi-scale contribution of copy number alterations to cancer subtypes and survival. 

\begin{figure}[ht]
    \centering
    \includegraphics[width=\textwidth]{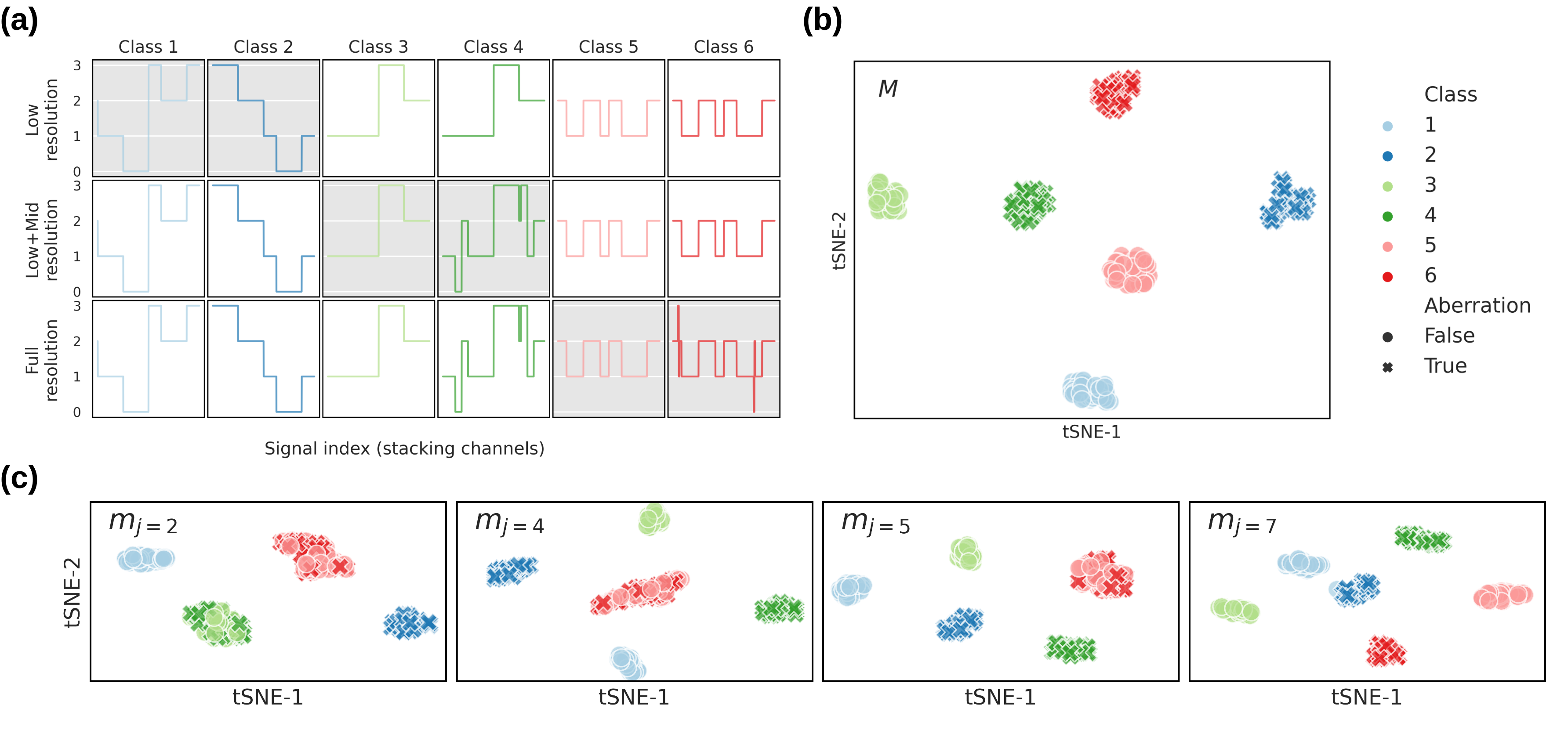}
    \caption{{\bf Illustrative Example}. Source-separated signals are adaptively filtered to obtain scale embeddings. These scale embeddings are then combined to obtain a single scale-attentive representation. (\textbf{a}) Simulated noise-free signals. As the scale increases, more pairs of classes become distinguishable (highlighted). Only the first pair of classes are distinguishable at the low-scale (top row), the second two at medium scale (middle row), whilst the final two become distinguishable only at high scales (bottom row). (\textbf{b}) tSNE projection of our multi-scale embedding ($\mathbf{M}$). (\textbf{c}) tSNE projection of our adaptively learnt scale specific embeddings ($\mathbf{m}_j$). }
         \label{fig:demo}
\end{figure}

{\bf Illustration.}~We first illustrate using a toy example in Figure \ref{fig:demo}a. Here, we simulated six signal classes which are displayed at low, mid and full scale (raw signal). Notice that at low scales, classes 3/4 and 5/6 are indistinguishable but, at higher scales, these classes can be separated using the fine scale signal features. Figure \ref{fig:demo}c shows a series of scale-dependent Wave-LSTM embeddings derived from the simulated data. The low-scale Wave-LSTM embeddings contain only four distinct groups corresponding to the four distinguishable signal classes. While the higher-scale Wave-LSTM embeddings contain increasing numbers of classes from 4 to 6. Figure \ref{fig:demo}b shows an overall \emph{multi-scale embedding} in which Wave-LSTM combines information across all scales. When applied to copy number data we expect this capability to allow us to separate the effects of large-scale chromosomal alterations from finer-scale, focal aberrations in unsupervised and supervised learning tasks.

{\bf Reproducibility.}~We make our code available as a PyTorch package with accompanying notebooks: \url{https://github.com/cwlgadd/waveLSTM}.

\section{Methods}
\begin{figure}[t]
    \centering
\includegraphics[width=\textwidth]{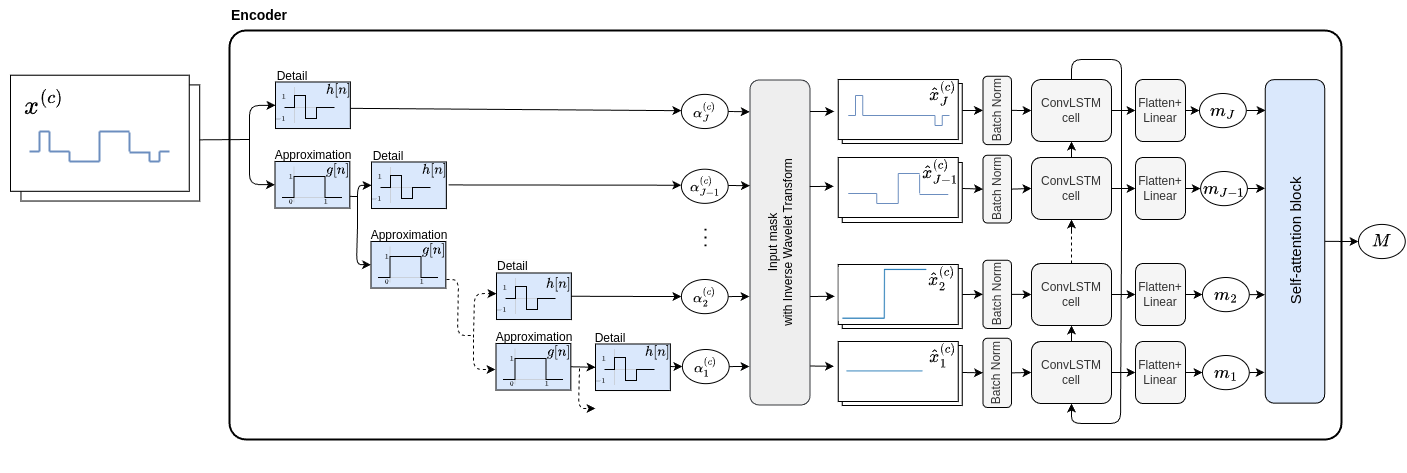}
    \caption{{\bf The Wave-LSTM Encoder}. Source separation is achieved through zero-masking of the multi-resolution wavelet cascading filter bank. This is then adaptively filtered and combined to output a scale-attentive encoding.}
    \label{fig:waveLSTM}
\end{figure}

{\bf Overview.}~Wave-LSTM consists of three main integrated components: (i) wavelet-based source separation, (ii) source-specific convolutional filters and (iii) a self-attention block that examines information  across scales. This leads to an output representation of the input signal which can be used directly for unsupervised analyses or integrated into a prediction/classification module for supervised learning (Figure~\ref{fig:waveLSTM}).

{\bf Notation.} We briefly define some notation that will be of use in the understanding of the next sections. Let $X_i\in\mathbb{R}^{C \times W}$ denote the $i$-th training sequence which, in generality could be composed of $C$ channels, each of width $W$. We index each channel as $x^{\left(c\right)} \in \mathbb{R}^W$, excluding sample index $i$ for notational clarity. We then denote by $n$ the position along each channel, and for brevity and notational convenience we assume channels are binned to a width that is shared across channels. 

In our application, each channel then corresponds to the division of the genome across chromosomes and strands (i.e. major and minor copy number). For the human genome, this then corresponds to $C=46$ channels from $23$ chromosomes (when including non-autosomes), across $2$ strands. 

Consequently, $x^{\left(c\right)}[n]$ then denotes the sequence of binned copy numbers at each (proportional) position along the respective chromosome and strand. We refer to these positions along a channel as the loci.

{\bf Wavelet-based source separation.} We use the discrete wavelet (Haar) transform as a cascading filter bank. Each level of the filter leads to a corresponding low-pass decomposition $u$ (which approximates the signal), and high-pass decomposition $w$ (which details the difference between the current level and the approximation), with impulse responses $u[\cdot]$ and $w[\cdot]$, respectively. Wavelet coefficients are obtained by repeatedly applying these low-pass, and high-pass filters along each channel $x^{\left(c\right)}[n]$, and then down-sampling. At the first level, this is then equivalent to performing the convolutions $
    y_{\text{high}}[n] \coloneqq (x^{\left(c\right)} * w)\downarrow Q = \sum_k x^{\left(c\right)}[k]w[Qn-k],$ and $
    y_{\text{low}}[n] \coloneqq (x^{\left(c\right)} * u)\downarrow Q = \sum_k x^{\left(c\right)}[k]u[Qn-k]$ 
where $Q$ is known as the decimation factor and $\downarrow$ is the subsampling operator.

This obtains detail coefficients $y_{\text{high}}[n] \coloneqq \coef^{\left(c\right)}_J$, and approximation coefficients $y_{\text{low}}[n]$.
As depicted in Figure~\ref{fig:waveLSTM}, we repeat this procedure on each subsequent set of approximation coefficients, at each stage obtaining the next set of detail coefficients, $\coef^{\left(c\right)}_{j}$, and a new set of approximation coefficients. This is performed independently across each input channel, obtaining a sequence of wavelet coefficients for each. This sequence of coefficients details information of increasing scale within a channel, and we index $j$ over these scales. 
We can use this sequence of coefficients to recover scale specific components of the original input signal using the Inverse Wavelet Transform (IWT). This is achieved by zero-masking the $\alpha^{\left(c\right)}_{\backslash j}$ components to obtain $\hat{x}^{\left(c\right)}_\j = \text{IWT}\bigl([\mathbf{0}, \dots, \mathbf{0}, \mathbf{\coef}^{\left(c\right)}_{\j}, \mathbf{0}, \dots, \mathbf{0}]\bigr)$. Repeating this for each channel at scale $j$, we denote $\hat{X}_j$ as the concatenation of channels $\hat{x}_\j^{\left(c\right)}$, and obtain a sequence of source separated, scale specific components \[\left\{\hat{X}_j\right\}_{j=1}^J \vcentcolon= \left\{\left[ \hat{x}_j^{\left(1\right)},\dots, \hat{x}_j^{\left(C\right)}\right]\right\}_{j=1}^J.\]
We can choose $J$ to be either the maximum depth, or a truncation of the sequence which then applies a de-noising effect.

An additional advantage of truncation is that it also applies a form of average-pooling on our input when components $j>J$ are excluded from the IWT~\citep{mallat1999wavelet}. Consequently, each source separated signal produced needs to only be long enough to capture the finest details, specified through the chosen wavelet's decomposition length and the choice of $J$. For example, given a wavelet with a decomposition length of $2$ and a recursion depth of $J=6$, we can accurately capture artifacts longer than $2^{-J}=1/64$-th of the input signal, with each source-separated input component being $64$ elements long.

{\bf Convolutional-LSTM cell.} Long Short-Term Memory (LSTM) networks are a type of recurrent artificial neural network which are specifically designed towards processing sequential data. \cite{shi2015convolutional} introduced a convolutional LSTM network for channelised signals, which is defined through a set of recurrence relations. We extend upon this by adding a convolutional \emph{recurrent projection layer} to the output gate. Our contribution then simply replaces the typical output gate with 
\begin{equation}
    H_\j = \tanh\left(W_{oh} * \left(o_\j  \odot \tanh(C_\j)\right)\right),
\end{equation}
where $C_j$ is the tensor of cell states composed of $h_c$ channels, $H_j$ is the tensor of hidden states now composed of $h_{\text{proj}} < h_c$ channels, $o_j$ is the response of the output gate, whilst $W_{oh}$ and convolutional operator $*$ define the recurrent convolutional projection. This is a modified version to the recurrent projection proposed by~\citet{sak2014long}, and reduces the number of channels of the hidden short-term memory state vector whilst maintaining the ability to stack LSTM layers. This was originally proposed for large vocabulary models which necessitated a larger long-term memory cell that led to prohibitive memory requirements. With this projection, the cell and hidden state dimensions are uncoupled and the flow of information from coarse-scale embeddings to fine-scale embeddings can be controlled. We couple this with a linear non-recurrent output projection to obtain the \emph{scale embedding}, $m_\j = W_{\j}F(H_\j)$ where $F\left(\cdot\right)$ denotes a flattening layer which concatenates each hidden channel and $\mathbf{m}_j \in \mathbb{R}^D$.

\begin{figure}[b]
     \centering
     \includegraphics[width=0.9\textwidth]{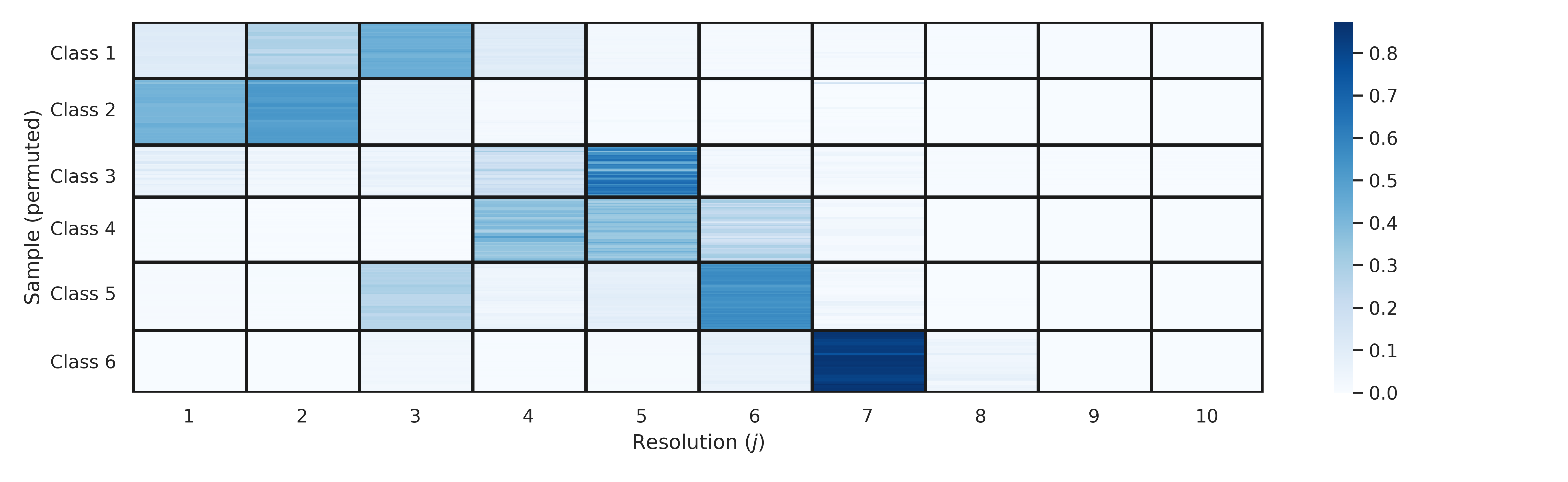}
         \caption{{\bf Self-attention of the Illustrative Example} ($\mathbf{A}\in\mathbb{R}^{1\times J}$), for each increasing scale. Samples are ordered by label, then permuted via spectral bi-clustering. We observe that attention is given to higher scales in the presence of transient signals such as finer-scale, focal aberrations.}
         \label{fig:sim_clf_atn}
\end{figure}

{\bf Self-attention block.} Having obtained a sequence of scale-specific embeddings, we now aim to find a single multi-scale representation that can be used for prediction tasks. Towards this end we employ the self-attentive embedding of~\citet{lin2017structured}. Originally developed to convert a sequence of word embeddings to a sentence embedding, we instead convert our sequence of \emph{scale embeddings} to a single \emph{multi-scale embedding}. We achieve this by
taking a linear combination of the $J$ LSTM hidden vectors in $\mathbf{S} = \left[\mathbf{m}_1^T, \dots, \mathbf{m}_J^T\right]$ of size $D$-by-$J$, using a self-attention mechanism. The attention mechanism takes the scale embeddings $\mathbf{S}$ as input, and outputs weights
\begin{equation}
    \mathbf{a} = \text{softmax} \left ( \mathbf{w}_{s2}\text{tanh} \left ( \mathbf{W}_{s1}\mathbf{S} \right ) \right ),
\end{equation}
where $\mathbf{W}_{s1}$ is a weight matrix of shape $d_a$-by-$D$, and $\mathbf{w}_{s2}$ is a row vector of parameters with size $1$-by-$d_a$, and we obtain an attention vector with $J$ elements, whose weights sum to one. This attention vector can explain a singular semantic meaning between different scales through $\mathbf{m}=\mathbf{a}\mathbf{S}^T$. For example, in some outcome prediction tasks in cancer genomics, the presence of whole genome doubling (at a coarsest scale) and a gene mutation (at a fine scale) may have some semantic meaning when occurring together. However, there may exist many meaningful combinations across other scales, and so we extend $\mathbf{w}_{s2}$ to an $r$-by-$d_a$ matrix $\mathbf{W}_{s2}$, where $r$ is the maximum number of semantic combinations we can model, and obtain the attention matrix
\begin{equation}
\label{eqn:self-attention:A}
    \mathbf{A} = \text{softmax}\left(\mathbf{W}_{s2}\text{tanh} \left ( \mathbf{W}_{s1}\mathbf{S} \right ) \right ).
\end{equation}

The \emph{multi-scale embedding} matrix for a single sample is then the $r$-by-$D$ matrix, $\mathbf{M} = \mathbf{A}\mathbf{S}^T$. To utilise this multi-scale embedding, we flatten each row of $\mathbf{M}$ and use this as input to a decoder/task-specific head, training end-to-end. This is demonstrated by our running example using a simple fully connected classifier (previously shown in Figure~\ref{fig:demo}). Figure~\ref{fig:sim_clf_atn} shows the corresponding attention matrix $\mathbf{A}$. Notice that attention is paid to resolution 7 for Class 6 which is only different from Class 5 in terms of fine detailed structures. Attention is paid to mid-resolution scales 4/5 for Classes 3 and 4 where mid-sized differences are presence. While attention is only needed to paid to resolutions 1-3 for classes 1 and 2 which differ by large, broad alterations.

where we correctly identify a multi-scale embedding that is attentive to the finer scales in the presence of aberrations and coarse scales otherwise, as shown in . 

\section{Experiments}
\begin{figure}[t]
	\centering
	\includegraphics[width=\textwidth]{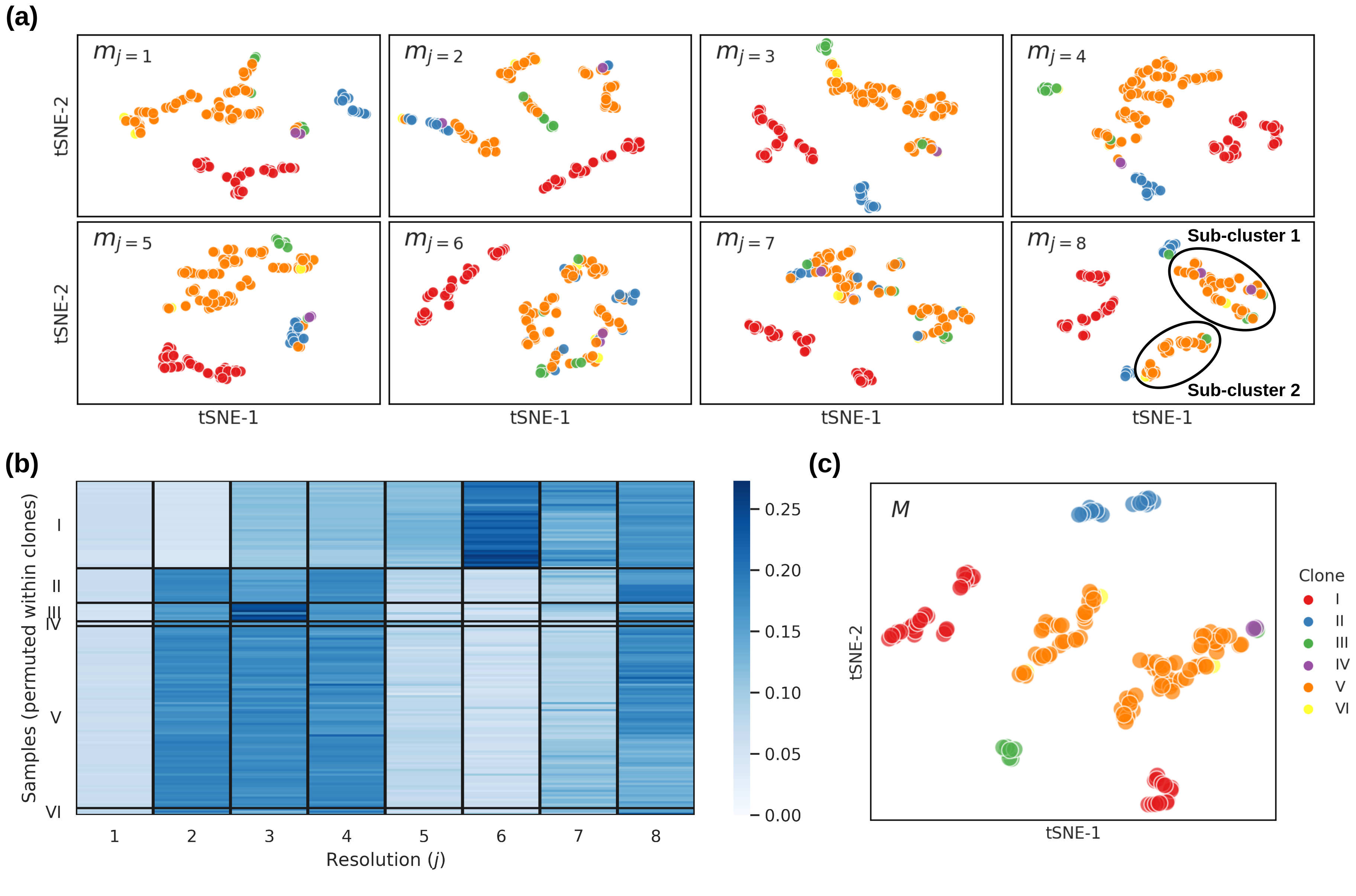}
	\caption{{\bf Single-cell copy number scale embeddings}. (\textbf{a}) t-SNE plots of scale embeddings ($\mathbf{m}_j\in\mathbb{R}^3$). (\textbf{b}) Attention ($\mathbf{A}\in\mathbb{R}^{1\times J}$), for each class and scale. (\textbf{c}) tSNE projection of multi-scale embedding ($\mathbf{M}\in\mathbb{R}^{1\times 3}$).}
	\label{fig:clonal_res}
\end{figure}

{\bf Single cell analysis.} We first consider the task of detecting clonal structures from single-cell profiling of copy number alterations. Cancers evolve through an iterative process whereby successive generations of cancer cells can accumulate somatic mutations. Over time, this process leads to related but genetically distinguishable descendant sub-populations within a single tumour. These are referred to as sub-clones, and their collective clonal structure reveals key aspects of tumour development which can be useful for targeted treatment and early intervention. 

We obtained a 10x Genomics Chromium single-cell DNA sequencing dataset which contains cells obtained from frozen breast tumour tissue from two patients, at five tumour sections and sequenced with a low coverage between 0.02x and 0.05x\footnote{URL: \url{https://github.com/raphael-group/chisel-data}}. However, for our experiments we follow previous work and consider only site E of patient S0, comprising of $2,075$ sequenced cells. \citet{zaccaria2021characterizing} used the CHISEL algorithm to allocate $1,448$ of the $2,075$ tumour cells to 6 clones, including one diploid clone (labelled I) comprising mostly normal cells, and 5 aneuploid clones which contain varying degrees of copy number alterations (labelled II-VI) (the remaining cells were unclassified and we excluded them from the analysis). 

Using the CHISEL labels as an \emph{assumed} soft ground-truth\footnote{There is no objective ground truth in such settings.}, we compared these labels to the embeddings obtained on an auto-encoding task using the \mname~encoder. A similar task was also performed by~\citet{yu2023rccae} using a Convolutional Neural Network (CNN) auto-encoder, and we use the same decoding architecture for comparison. Following~\citet{yu2023rccae}, we encode to $3$-dimensional embedding spaces, s.t. $\mathbf{m}_j\in\mathbb{R}^{D=3}$, and use a single attention hop, s.t. $\mathbf{M}\in\mathbb{R}^{1\times3}$. Additionally, we choose to stack chromosomes such that $W=256\times22$ (using only autosomes), $C=2$, and choose $J=8$. This is both in keeping with~\citet{yu2023rccae}, and allows us to view coarser scale embeddings.

\begin{figure}[t]
    \centering
    \includegraphics[width=\textwidth]{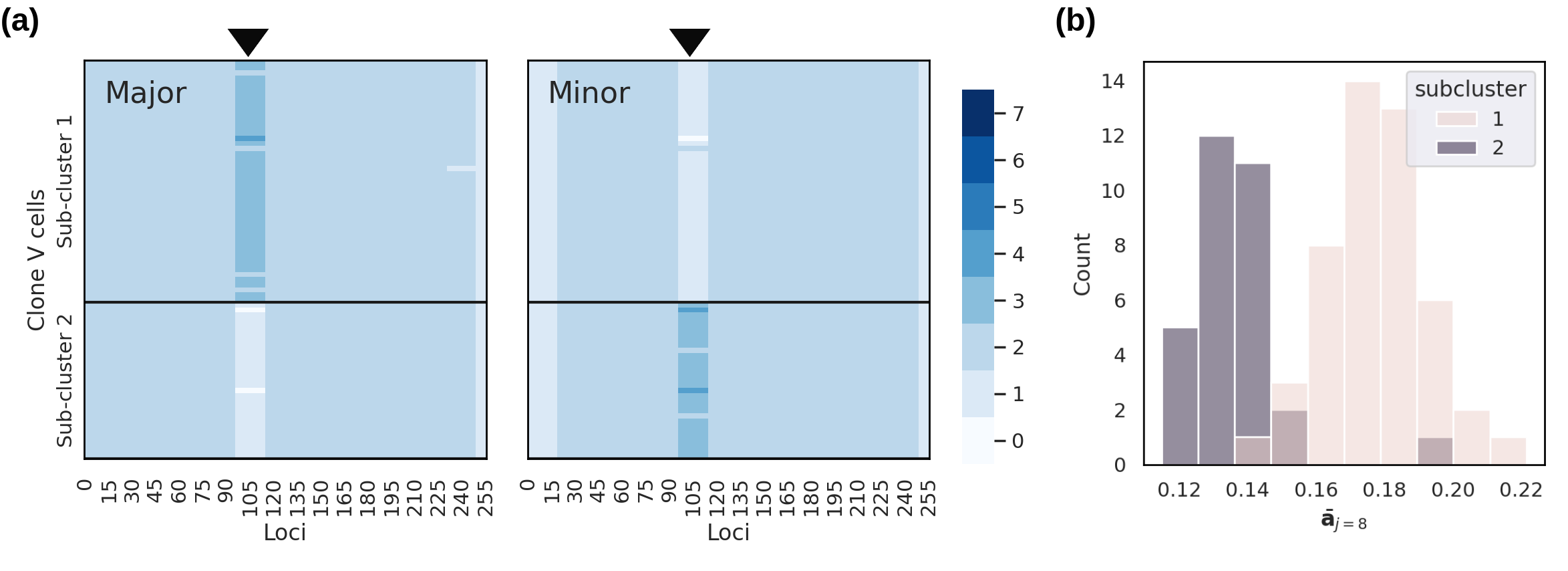}
    \caption{{\bf Novel sub-clone discovery}. (\textbf{a}) Copy number profile of clone V's major (left) and minor (right) strands of chromosome $1$. Samples are split into two sub-clusters in the embedding $h_{j=8}$ which are segregated by a systematic copy number difference at a single locus indicated by the black arrows. (\textbf{b}) The amount of attention placed on the $j=8$ scale differs for each sub-cluster.}
    \label{fig:clonalV_separation}
\end{figure}

Two-dimensional t-SNE plots of the three-dimensional scale and multi-scale embeddings are shown in Figure~\ref{fig:clonal_res}a and~\ref{fig:clonal_res}c, respectively, coloured by CHISEL labels. Scales from $j=1$ to $j=5$ showed latent structures (clusters) that are broadly consistent with the CHISEL clonal structure but sub-structures not previously captured by CHISEL or rcCAE also emerge. For example, consider clone V at $j=8$ which is split into two groups. In Figure~\ref{fig:clonalV_separation}a, we observe that this is due to the cells in clone V containing a small copy number gain on opposite strands. This separation is also reflected in the self-attention mechanism, shown in Figure~\ref{fig:clonal_res}b where cell samples are permuted through clustering of the attention matrix, and in Figure~\ref{fig:clonalV_separation}b where we see each sub-cluster places a different level of attention on this scale. This is also clearly demonstrated in clone II at scale $j=6$, which is grouped into two clusters. Upon inspection, this distinction is based on deletions on chromosomes $6, 17$ and $22$.

Overall, \mname~is able to recover clonal patterns from single-cell sequencing data in an unsupervised fashion that are broadly consistent with current state-of-the-art methods such as CHISEL. However, it also provides enhanced exploratory capabilities to look at multi-scale structures which identified novel lineages of cancer cells.

{\bf Cancer survival prediction.}  Copy number alterations are known to be predictive of survival outcome in cancer \citep{hieronymus2018tumor}. In this section we consider the use of \mname~as a means of providing multi-scale information from copy number profiles, with the goal of predicting cancer survival outcomes. 

The inclusion of CNAs in survival modelling can be challenging due to the multi-channeled, spatially correlated sequence data. We used \mname~to encode the CNA profiles as an input into a deep survival model, DeSurv \citep{danks2022derivative}, alongside other features, to predict survival outcomes. Note that the choice of DeSurv is for convenience as it is a flexible neural network based approach for modelling survival distributions in continuous time. Cox-based survival models (e.g. PyCox \citep{kvamme2019time}) or alternative neural approaches (e.g. SumoNet \citep{rindt2022survival}, Neural Fine-Gray \citep{jeanselme2023neural}) could also be used.

{\bf Simulation study.} We simulate two cancer types, each described by a different underlying  linear Cox proportional hazard models with Gompertz distributed survival functions. Consequently survival times are generated by 
\[
T = \frac{1}{\beta} \ln\left(1 - \frac{\beta \ln(u)}{\alpha\exp\{\sum_{i} X_i\}}\right),
\]

where $u$ is a random variable uniformly distributed on the unit interval, and $\{X_i\}_{i=1}^{12}$ is the sequence of covariates, including standard Gaussian age, Bernoulli distributed gender, and $10$ uniformly distributed insertion locations of equal width across $2$ channels. Using this model we sample survival outcomes originating from two cancer types, described by $\alpha=0.1$ and $1.5$ respectively. For each, the rate of censoring was kept equal and we choose $\beta=1$ .
For us to test the capacity of \mname~ to identify small aberrations, the CNA profiles from one model were given an additional CNA gain event distinguishing it from the other cancer type. These were placed in the same position, at $1/64$ of a channel width.
Consequently, the only information which then distinguishes cancer types is this additional gain event.

\begin{table}[t]
    \centering
    \begin{adjustbox}{width=0.65\textwidth} 
    \begin{tabular}{c||c||c|c|c}
        Study & Encoder & $C_{td} (\uparrow)$ & IBS $(\downarrow)$ & INBLL $(\downarrow)$  \\ \hline \hline
        \multirow{4}{*}{\rotatebox[origin=c]{90}{Simulation}} & Average CN & $ 0.54\pm 0.004$ & $0.16\pm0.002 $ & $0.49\pm0.004$  \\ \cline{2-5}
        & rcCAE & $0.71\pm0.003$ & $0.06\pm0.003$ & $0.23\pm0.040$  \\ \cline{2-5}
        &LSTM & $0.68\pm0.002$ & $0.06\pm0.004$ & $0.22\pm0.010$  \\ \cline{2-5}  
        &\textbf{Wave-LSTM} & ${\bf 0.72 \pm 0.005}$ & ${\bf 0.04 \pm 0.006 }$ & ${\bf 0.14 \pm 0.017}$ \\ \hline\hline
        \multirow{4}{*}{\rotatebox[origin=c]{90}{TCGA}} & Average CN  & $0.62\pm 0.002$ & $0.13\pm 0.002$ & $0.40\pm 0.005$  \\ \cline{2-5}
        & rcCAE & $0.75\pm 0.011 $ & $0.11 \pm 0.007 $ & $0.34\pm 0.016$  \\ \cline{2-5}
        &LSTM & $0.72 \pm 0.007$ & $0.11\pm0.005 $ & $0.34 \pm 0.015$  \\ \cline{2-5}  
        &\textbf{Wave-LSTM} & ${\bf 0.78 \pm 0.008}$ & ${\bf 0.10 \pm 0.010}$ & ${\bf 0.31 \pm 0.024}$  \\ \hline
    \end{tabular}
    \end{adjustbox}
    \vspace{0.1cm}
    \caption{{\bf Survival performance metrics}, with one standard deviation across five random seeds.}
    \label{tab:ascat_surv}
\end{table}

We compared predictive performance to three baselines. The first was a simple baseline summary statistic, the genome-wide average copy number, which has been historically used as a measure of overall tumour genomic complexity (Average CN). Additionally, we also compared to a CNN encoder architecture of~\cite{yu2023rccae} (rcCAE), and an LSTM-based encoder in which recurrence is over loci. These alternative baselines account for the sequential and spatially correlated nature of the data but not the multi-scale aspect. All use the same DeSurv survival model. We again reiterate and emphasise that  we are unaware of \emph{any} previously developed representation learning approach for copy number data in the literature. Hence, we created these baselines to test the added benefit of the wavelet and attention components of Wave-LSTM over these simpler encoding mechanisms.

We report three metrics: Time-dependent Concordance Index ($C_{td}$, higher is better) measures how well a model distinguishes risk between samples at different times. The Integrated Brier Score (IBS, lower is better) measures if the model can accurately predict the probability of an event occurring over a period, whilst the Integrated Negative Binomial Log-Likelihood (INBLL, lower is better) measures how well the model prediction matches the actual occurrence of the event. 

Table \ref{tab:ascat_surv} shows that the predictive performance of \mname~exceeds all other baselines on all three measures. In particular, the IBS and INBLL measures were significantly improved, indicating that the calibration of \mname~predictions was superior.

{\bf The Cancer Genome Atlas} We next obtained copy number data from multiple cohorts from The Cancer Genome Atlas (TCGA)~\citep{weinstein_cancer_2013} to test the utility of multi-scale representations in improving survival prediction accuracy. Specifically, we consider the survival outcomes of patients with Thyroid Cancer (THCA), Breast Cancer (BRCA), Ovarian cancer (OV), Glioblastoma Multiforme (GBM), and Head and Neck Squamous cell Carcinomas (HNSC), listed in order of typically decreasing expected survival outcome. In total, we examined $3021$ samples. This selection was used as a control to determine if \mname~was able to recover latent quantities reflecting this survival heterogeneiety. Copy number profiles were pre-computed using Allele-Specific Copy number Analysis of Tumors (ASCAT) ~\citep{van2010allele}\footnote{URL: \url{https://github.com/VanLoo-lab/ascat/tree/master/ReleasedData/TCGA_SNP6_hg19}}. In this example we channelise chromosomes.

\begin{figure}[!t]
    \centering
    \includegraphics[width=0.5\textwidth]{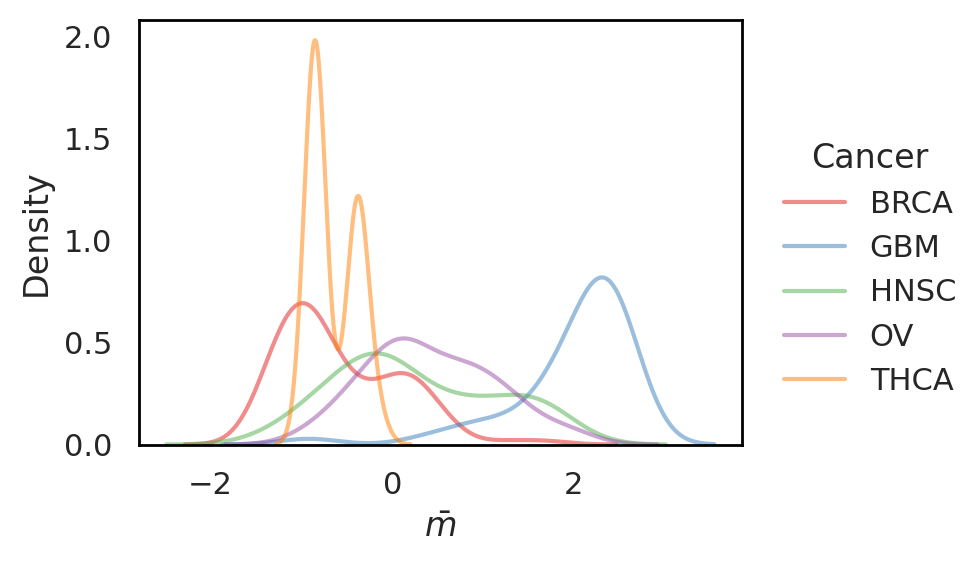}
    \caption{{\bf The Cancer Genome Atlas Survival analysis}. Wave-LSTM maps different cancer types along the multi-scale embedding dimension ($\bar{\mathbf{m}}$) in order of expected survival rates. Note that some cancer types (e.g. thyroid cancer, THCA; breast cancer, BRCA) appear multimodal in distribution which corresponds to different molecular subtypes within the respective cancer type.}
    \label{fig:ascat-surv-mres}
\end{figure}

The learnt multi-scale representations are plotted in Figure~\ref{fig:ascat-surv-mres} using kernel density estimation and grouped by cancer type. We observe increasing values correspond to worsening survival outcome showing that \mname~was able to identify a latent dimension associated with cancer survival. Additionally, we find two modes to thyroid cancer, and upon inspection we identified that Wave-LSTM had identified the previously known prognostic significance of male gender which is associated with poorer survival outcomes in papillary thyroid cancer than females \citep{liu2017reevaluating}.

We again compared our approach to alternative encoder baselines in Table~\ref{tab:ascat_surv}, where the \mname~encoder significantly outperforms the competing encoding approaches on all scores, suggesting that its specific design is encapsulating more information from the data.

\section{Discussion}
Within this work we introduced \mname, a multi-scale signal representation learning module, motivated by the analysis of cancer sequencing data in which mutations may occur across entire chromosomal regions, or isolated to a small number of genes. \mname~decomposes signals into multiple frequency bands, producing a series of band-pass outputs which are able to represent local transient structures at different scales. For each, we sequentially learn a scale-specific embedding using a novel LSTM framework. These are then combined using a self-attention mechanism to obtain a multi-scale embedded representation which can then be used in downstream tasks. 

Alongside a demonstrative classification example, two examples analysing cancer sequencing data are presented. We first validated \mname's ability to recover genetically distinguishable descendant sub-populations, recovering clonal patterns that are consistent with state-of-the-art specialist bioinformatics methods, whilst providing enhanced exploratory capabilities which identified novel lineages of cancer cells. Second, we used \mname~to learn multi-scale information of copy number profiles with the goal of predicting cancer survival times. We demonstrated significant improvements on a number of benchmark metrics against competing methods.

Alternatives to the LSTM architecture could be employed, such as Transformers \citep{vaswani2017attention} or more recently introduced alternatives such as selective state-space models \citep{gu2023mamba}. These architectures have demonstrated powerful capabilities for a number of applications and data types. However, the complexity of these models relative to our LSTM architecture may preclude their use in genomic data settings where sample sizes are substantially more limited than in natural language, image and video data applications. Wave-LSTM can also be generally applied to obtain multi-scale embeddings of complex structured, high-dimensional multi-channel input objects. We have tested its utility for genomic copy number analysis but its wider utility for other data types has not been examined here. 

\section*{Acknowledgements}

CG and CY are supported by an EPSRC Turing AI Acceleration Fellowship (Grant Ref: EP/V023233/1).

\pagebreak
\bibliographystyle{dinat}    
\bibliography{references}  

\end{document}